\documentclass[aps,showpacs,preprintnumbers,amsmath,amssymb]{revtex4}

\UseRawInputEncoding
\usepackage{graphicx}
\usepackage{epstopdf}
\usepackage{color}
\usepackage{subfig}
\usepackage{aasmacros}
\usepackage{amsmath,amssymb,amsfonts}

\begin{document}
 \captionsetup{justification=raggedright,singlelinecheck=false}
    \baselineskip=0.8 cm
    \title{\bf Thick accretion disk configurations around a compact object in the brane-world scenario}

    \author{Yunzhu Wei$^{1}$,
    Songbai Chen$^{1,2}$\footnote{Corresponding author: csb3752@hunnu.edu.cn},
    Jiliang Jing$^{1,2}$ \footnote{jljing@hunnu.edu.cn}}
    \affiliation{$^1$Department of Physics,  Key Laboratory of Low Dimensional Quantum Structures
    and Quantum Control of Ministry of Education, Institute of Interdisciplinary Studies, Synergetic Innovation Center for Quantum Effects and Applications, Hunan
    Normal University,  Changsha, Hunan 410081, People's Republic of China
    \\
    $ ^2$Center for Gravitation and Cosmology, College of Physical Science and Technology, Yangzhou University, Yangzhou 225009, People's Republic of China}

    \begin{abstract}
    \baselineskip=0.6 cm
    \begin{center}
    {\bf Abstract}
    \end{center}

   We have studied equipotential surfaces of a thick accretion disk around a Casadio-Fabbri-Mazzacurati (CFM) compact object in the brane-world scenario, which owns a mass  parameter together with a parameterized post-newtonian (PPN) parameter. With the increase of the PPN parameter, the size of the thick accretion disk decreases, but the corresponding Roche lobe size increases. Thus, the larger PPN parameter yields the larger region of existing bound disk structures where the fluid is not accreted into the central wormhole. Moreover, with the increase of the PPN parameter, the position of the Roche lobe gradually moves away from the central compact object, the thickness of the region enclosed by the Roche lobe decreases near the compact object, but increases in the region far from the compact object. Our results also show that the pressure gradient in the disk decreases with the PPN parameter. These effects of the PPN parameter on thick accretion disk could help to further understand compact objects in brane-world scenario.

    \end{abstract}

    \maketitle
    \newpage

    \section{Introduction}

    A compact object at the center of galaxy should own an accretion disk in which matter flows spiral into the central celestial body together with dropping off their initial angular momentums outwards as well as releasing their gravitational potential energy into heat.  The analyses of the properties of accretion disks can disclose the motion of matter near the central celestial body, which could offer the opportunity to capture some information from the celestial body because the accretion occurs in the strong gravity region. Therefore, the studies of accretion disks have contributed to examining the predictions from various theories of gravity including general relativity, which may deepen the understanding of the gravitational interaction \cite{1973A&A....24..337S,10.1111/j.1365-8711.1998.01774.x,1990MNRAS.244..377S,1975ApJ...199L.153E,1995ApJ...452..710N,Turolla_2000,
    Davis_2006,1977ApJ...214..840I,2002MNRAS.334..383F,1982ApJ...253..897P,2012arXiv1203.6851M,PMID:28179840,Liu:2022cph,
    Bahamonde:2015uwa,Ahmed:2015tyi,Jusufi:2021lei,Chen:2022scf,Liu:2022ruc,Zhou:2021cef}.
   Generally, in term of their geometrical thickness, the accretion disk models can be classified into two types. One of them is the so-called geometrically thin model in which the disk height is much smaller than the characteristic radius of the disk \cite{1973A&A....24..337S,10.1111/j.1365-8711.1998.01774.x,1990MNRAS.244..377S}. The heat generated by stress and dynamic friction in the disk can be effectively dispersed through the radiation over its surface, which leads to that the disk is cool. The other type is the geometrically thick model \cite{1975ApJ...199L.153E,1995ApJ...452..710N,Turolla_2000,Davis_2006,1977ApJ...214..840I,2002MNRAS.334..383F,1982ApJ...253..897P} in which the energy conversion into radiation is inefficient and the temperature of the accretion gas is higher than that in the previous thin disk model. It is widely believed that there exist thick accretion disks in the vicinity of many X-ray binaries and active galactic nuclei.

In the real astrophysical systems, the matter accretion is a highly complicated dynamic process and its complete description must resort to the high precise numerical calculations, such as general relativistic magnetohydrodynamics (GRMHD) simulations. However, in the past few decades, a simple and analytical model of geometrically thick and stationary tori orbiting black holes, known as Polish doughnuts \cite{1982ApJ...253..897P}, has been attracted a lot of attention. Although  the matter in this model is assumed to be in equilibrium and is no actually accreted by the black hole, the configurations of these geometrically thick equilibrium tori carry a lot of important characteristic information on the spacetime in the strong field regions.  Moreover, due to the fluid being in equilibrium, Polish doughnuts are often
used as an initial condition for numerical simulations of accretion flows. Thus,
the equilibrium tori around black holes have been studied in spacetimes in
general relativity and in other alternative theories of gravity (for a review see e.g. \cite{2012arXiv1203.6851M,PMID:28179840,Liu:2022cph}). Recently, thick accretion disks have been investigated in the background of the spherically symmetric black hole in Born-Infeld teleparallel gravity \cite{Bahamonde:2021srr} and probe effects of the teleparallel parameter on the equilibrium tori around the black hole, which show that the size of the disk  monotonically decreases with the teleparallel parameter \cite{2022PhRvD.106h4046B}.
 The non-selfgravitating equilibrium tori have also been studied in the background of the parameterised Rezzolla-Zhidenko black hole \cite{Cassing:2023bpt}. It is found that there exist standard ``single-torus" and non-standard ``double-tori" solutions within the allowed space of parameters, which means that the parameterised Rezzolla-Zhidenko black hole owns a much richer class of equilibrium tori. Moreover, the magnetized accretion disks around Kerr black holes with scalar hair have been respectively studied with  the constant
angular momentum  \cite{Gimeno-Soler:2018pjd} and  the nonconstant angular momentum  \cite{Gimeno-Soler:2021ifv}, which could help further constrain the no-hair hypothesis by combining with future observations. The stationarily and geometrically thick tori with the constant angular momentum are researched in the background of a nonrotating black
hole in $f(R)$-gravity with a Yukawa-like modification to the Newtonian potential\cite{Cruz-Osorio:2021gnz}. Making a comparison with the Kerr black hole in the general relativity, it is easy to find that there are notable changes in the configurations of the disks. Moreover, the equilibrium solutions of magnetised, geometrically thick accretion disks are also studied with the nonconstant specific angular momentum distribution in the  Kerr black hole spacetime \cite{Gimeno-Soler:2023anr}.

We here focus on thick accretion disk configurations around a compact object in the brane-world theory. According to the brane-world scenario \cite{PhysRevLett.83.3370,PhysRevLett.83.4690}, the usual four-dimensional spacetime  might be a three-brane embedded in a five-dimensional spacetime (the bulk). All of matter fields including electromagnetic field are confined to the three brane, and only gravity can
freely propagate both in brane and bulk. High energy corrections and Weyl stresses from bulk gravitons mean that
a static black hole solution on the brane is no longer the Schwarzschild
solution \cite{Dadhich:2000am}. However, the Einstein field equations in five dimensions are found to admit more spherically
symmetric solutions on the brane than in four-dimensional general relativity. The first black hole solution on the brane obtained in \cite{Dadhich:2000am}, has the same form as the usual Reissner-Nordstr\"{o}m solution, in which a tidal Weyl parameter plays the role of the electric charge. The star
solution with a constant density interior has been studied on the brane \cite{Germani:2001du}. The black hole solutions in the brane-world model and the corresponding observable effects have also been widely studied \cite{2008CQGra..25g5012P,
2016PhRvD..93l4068M,2007PhRvD..76f6002H,1999PhRvL..83.3370R,2022arXiv220809589L,2022arXiv220315785T,PhysRevD.69.064020,PhysRevD.70.024010,HARKO2005471,
Harko_2006}.  The properties of a thin accretion disks around a brane world black hole have been investigated \cite{PhysRevD.78.084015}, and it is shown that the particular signatures appeared in the electromagnetic spectrum could offer a  the possibility to directly test physical models with extra dimension by
using astrophysical observations from accretion disks. We here will consider a spherically symmetric  solution in the brane world obtained by Casadio, Fabbri and Mazzacurati \cite{PhysRevD.65.084040}. The properties of a thin accretion disks around the CFM compact object have also been studied in \cite{PhysRevD.78.084015}. However, the thick accretion disk configurations around the CFM compact object remain open. The main purpose of this paper is to probe the properties of thick accretion disk configurations around the CFM wormhole. Actually, Casadio, Fabbri and Mazzacurati \cite{PhysRevD.65.084040} obtained
two  analytical solutions of the spherically symmetric vacuum brane-world, which are parameterized by the Arnowitt-Deser-Misner (ADM)
mass and the PPN parameters. The first solution is given by $g_{tt}=1-\frac{2M}{r}$ and $g_{rr}=\frac{r(2r-3M)}{(r-2M)[2r-M(4\beta-1)]}$,  where $\beta$ is a PPN parameter. After a careful analysis, one can find that the PPN parameter $\beta$ does not affect on the thick accretion disk configurations because the potential $W$  (which determines the
equipotential surfaces topology of the disk)  does not depend on the metric component $g_{rr}$. This means that the thick accretion disk configurations are the same as in the usual Schwarzschild black hole spacetime. Therefore, we here only consider the second  CFM brane-world  solution and study effects of PPN parameter on the configurations of the thick accretion disk.

The paper is organized as follows. In Sec. II, we will briefly review the second  CFM brane-world  solution \cite{PhysRevD.65.084040} and then analyze the changes of the marginally stable orbit and the marginally bound orbit with the PPN parameter for a timelike particle.
 In Sec. III,  we will investigate thick accretion disk configurations around the CFM brane-world compact object and probe effects of the PPN parameter on the disk configurations. Finally, we present a summary.

\section{Particle motions in the background of a compact object in the brane-world scenario}

Lets us now briefly review the second CFM brane-world solution in \cite{PhysRevD.65.084040} and its metric form is
\begin{equation}
        \label{metric1}
        {\rm d}s^2=-\left(1-\frac{1}{\gamma}+\frac{1}{\gamma}\sqrt{1-\frac{2\gamma M}{r}} \right)^2{\rm d}t^2+(1-\frac{2\gamma M}{r})^{-1}{\rm d}r^2+ r^2{\rm d}\theta^2 +r^2\sin^2\theta {\rm d}\phi^2,
\end{equation}
which is spherically symmetric since it is invariant under a rotation or reflection transformation.
The solution is asymptotically flat and owns an ADM mass  parameter $M$ and a PPN parameter $\gamma$. However, the geometric properties of the spacetime (\ref{metric1}) depend on the value of $\gamma$. As $\gamma=1$, it reduces to the usual Schwarzschild black hole spacetime and the event horizon is located at $r=2M$. As $\gamma>1$, one can find that the only a singularity in the metric lies at $r=r_0=2M\gamma$, where all the curvature invariants are regular. Moreover,  $r=r_0$ is a turning point for all physical curves. Thus, the metric (\ref{metric1}) describes geometry of a wormhole with a throat radius $r_{\rm throat}=2M\gamma$  in this case. As $\gamma<1$, the metric is singular at $r_0$ and at the  null surface $r=r_{\rm s}=2M/ (2-\gamma)$. Along this null surface the Ricci scalar $R$ diverges as $R\sim 1/(\sqrt{r-r_0}-\sqrt{r_H-r_0})$.  In other words, the metric (\ref{metric1}) describes geometry of a naked singularity with a singular null surface $r_{\rm s}=2M/ (2-\gamma)$. Therefore, the value of $\gamma$ plays a key role in the global causal structure of the spacetime (\ref{metric1}). The analyses of the structure spacetime shows that  the solution (\ref{metric1}) describes  geometry of a
pathological naked singularity as $ \gamma<1$, or a black hole as $\gamma=1$,  or a regular wormhole as $\gamma>1$ \cite{PhysRevD.65.084040,Tan:2020hog}.
\begin{figure}
    \includegraphics[width=4cm]{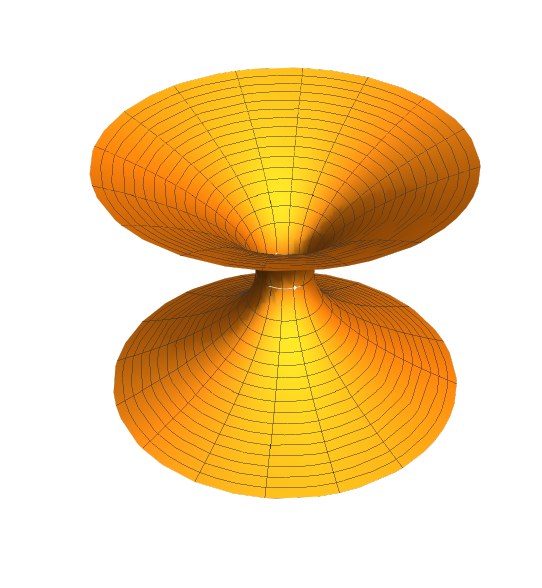}\includegraphics[width=4cm]{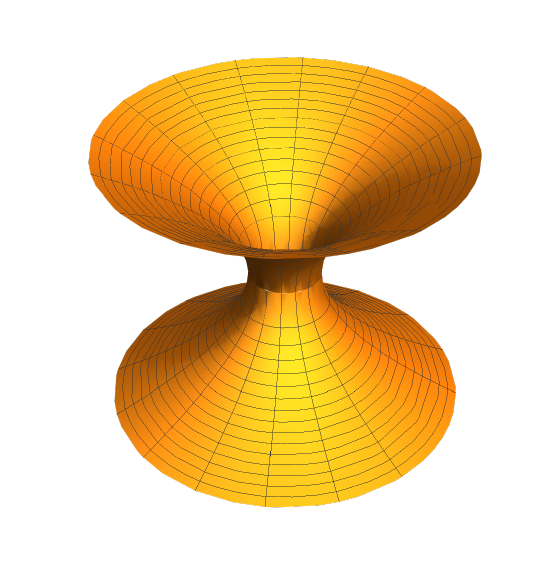}\includegraphics[width=4cm]{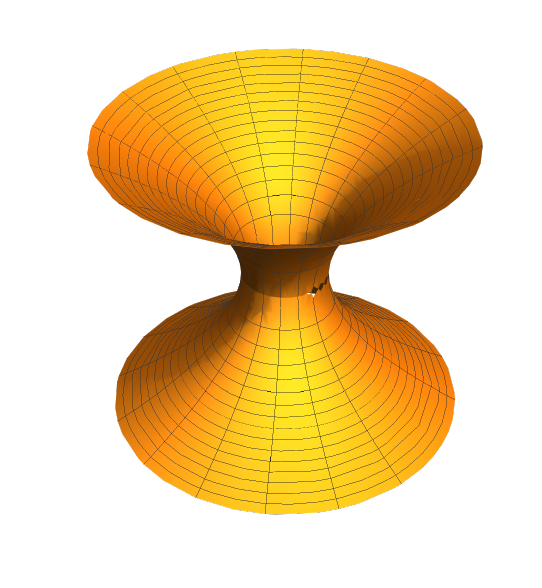}
    \caption{ The embedding diagrams of the metric (\ref{metric1}). The left, middle and right panels correspond to the cases $\gamma=0.8$, $1$ and $1.2$, respectively. Here we set $M=1$.}
    \label{diagram}
   \end{figure}
As in \cite{Jusufi:2021lei}, in order to visualize the spacetime (\ref{metric1}), we present the embedding diagrams in Fig.\ref{diagram} where the  equatorial slice $\theta=\pi/2$ at some fix moment in time $t=$constant is embed into three-dimensional Euclidean
space $ds^2=dz^2+dr^2+r^2d\phi^2$ with $dz=\pm \sqrt{r/(r-2M\gamma)-1}dr$.

To study the thick accretion disk configurations in the background of a compact object in the brane-world scenario (\ref{metric1}), one must get the radius of the marginally stable orbit $r_{\rm ms}$ and the marginally bound orbit $r_{\rm mb}$ for a single timelike particle moving in the spacetime, which are two essential quantities for determining thick disk configurations around a compact object. For the spacetime (\ref{metric1}), the Lagrangian density of a single timelike particle's motion
       \begin{equation}
        \label{lagelangri}
        \mathcal{L} = \frac{1}{2} g_{\mu\nu}(x)\dot x^\mu \dot x^\nu,
    \end{equation}
does not contain the time coordinate $t$ and the angular coordinate $\phi$, so there are two conserved quantities $E$ and $L$ for the particle, which respectively correspond to its energy and angular momentum. With these conserved quantities, the motion equation of the timelike particle moving in the equatorial plane can be further expressed as
    \begin{equation}
        \label{radialequation}
       \dot{r}^2-\left(1-\frac{1}{\gamma}+\frac{1}{\gamma}\sqrt{1-\frac{2\gamma M}{r}} \right)^{-2}\bigg(1-\frac{2\gamma M}{r}\bigg)\bigg[E^2-V_{\rm eff}(r)\bigg]=0,
    \end{equation}
   with the effective potential
   \begin{equation}\label{effpotenial}
    V_{\rm eff}(r)=\left(\frac{L^2}{r^2}+1\right) \left(1-\frac{1}{\gamma}+\frac{1}{\gamma}\sqrt{1-\frac{2\gamma M}{r}} \right)^2.
    \end{equation}
   From the condition of  circular orbit $V_{\rm eff}=E^2$ and $V'_{\rm eff}=0$ \cite{universe7010002,2021EPJC...81..875S,2022PhRvD.106h4046B}, one can obtain that
    \begin{eqnarray}\label{ELS}
    &&E^2=\frac{r-2\gamma M+(\gamma-1)\sqrt{r(r-2\gamma M)}}{r-3\gamma M+(\gamma-1)\sqrt{r(r-2\gamma M)}}\left(1-\frac{1}{\gamma}+\frac{1}{\gamma}\sqrt{1-\frac{2\gamma M}{r}} \right)^2,
    \nonumber\\
    &&L^2=\frac{\gamma M r^2}{r-3\gamma M+(\gamma-1)\sqrt{r(r-2\gamma M)}}.
    \end{eqnarray}
   Combining $E$ and $L$ in Eq.(\ref{ELS}) with  the condition $V''_{\rm eff}(r)=0$, one can find that
    \begin{equation}
   (\gamma^2-2\gamma+2)r^2+\gamma(3\gamma^2-6\gamma+11)Mr+12 M^2 \gamma^2+(\gamma-1)(2r-9M\gamma)\sqrt{r^2-2\gamma Mr}=0,
    \end{equation}
    which gives the radius of the marginally stable orbit
    \begin{eqnarray}\label{rms}
   r_{\rm ms}=\left\{\begin{array}{ll}
   \frac{M}{6(\gamma-2)}[4(3\gamma^2-6\gamma-4)-(1+\sqrt{3}\;{\rm i})\mathcal{A}_1\mathcal{B}^{-1/3}_1
   -(1-\sqrt{3}\; {\rm i})\mathcal{B}^{1/3}_1],&\quad\quad 0<\gamma\leq2,\\
    \frac{M}{3(\gamma-2)}[2(3\gamma^2-6\gamma-4)+\mathcal{A}_1\mathcal{B}^{-1/3}_1
   +\mathcal{B}^{1/3}_1],&\quad\quad \gamma>2,
     \end{array}\right.
    \end{eqnarray}
    with
    \begin{eqnarray}
   \mathcal{A}_1&=&3\gamma(\gamma-2)(3\gamma^2-6\gamma+19)+64,\nonumber\\
   \mathcal{B}_1&=&27\gamma(\gamma-1)(\gamma-2)\sqrt{-(9\gamma^4-36\gamma^3+93\gamma^2-114\gamma+176)}-27\gamma^6+162\gamma^5-459\gamma^4 \nonumber\\
    &+&756\gamma^3-1224\gamma^2+1368 \gamma-512.\nonumber
   \end{eqnarray}
  The marginally bound orbit $_{rmb}$ is the innermost unstable circular orbit for a timelike particle \cite{universe7010002,2021EPJC...81..875S,2022PhRvD.106h4046B}, which can be determined by $V_{\rm eff}=1$ and $V'_{\rm eff}=0$, i.e,
  \begin{equation}
   2(\gamma^2-3\gamma+2)r^2-\gamma(3\gamma^2-12\gamma+10)Mr+4 M^2 \gamma^2-2(\gamma-1)[(\gamma-2)r+3M\gamma]\sqrt{r^2-2\gamma Mr}=0.
    \end{equation}
   \begin{figure}
    \includegraphics[width=6cm]{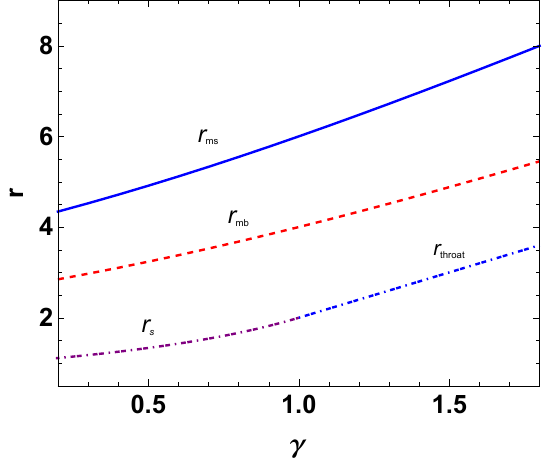}
    \caption{Changes of the marginally stable orbit radius $r_{\rm ms}$ and the marginally bound orbit radius $r_{\rm mb}$ with the PPN parameter $\gamma$. The blue line and the dashed red line denote the radii $r_{\rm ms}$ and $r_{\rm mb}$, respectively. The purple dot dash line corresponds to the position of the naked singularity $r_{\rm s}$ and the blue dot dash line is the throat radius $r_{\rm throat}$. Here we set $M=1$.}
    \label{frms}
   \end{figure}
   Solving the above equation, we obtain the radius of the marginally bound orbit
    \begin{eqnarray}\label{rmb}
   r_{\rm mb}=\left\{\begin{array}{ll}
   \frac{M}{12(\gamma-1)(\gamma-2)}[(9\gamma^3-24\gamma^2-8\gamma+24)+\mathcal{A}_2 \mathcal{B}^{-1/3}_2
  +\mathcal{B}^{1/3}_2],&0<\gamma\leq 0.94026, \quad or \quad \gamma >2, \\
    \frac{M}{24(\gamma-1)(\gamma-2)}[2(9\gamma^3-24\gamma^2-8\gamma+24)-\frac{(1+\sqrt{3}\;{\rm i})\mathcal{A}_2}{\mathcal{B}^{1/3}_2}
   -(1-\sqrt{3}\;{\rm i})\mathcal{B}^{1/3}_2],&\quad\quad 0.94026<\gamma\leq2,
     \end{array}\right.
    \end{eqnarray}
    with
    \begin{eqnarray}
   \mathcal{A}_2&=& 81\gamma^6-432\gamma^5+1008\gamma^4-1488\gamma^3+1696\gamma^2-1248\gamma+384,\nonumber\\
   \mathcal{B}_2&=&48(\gamma-1)^2 (9\gamma^3-32\gamma^2+32\gamma-8)
   \sqrt{-3(27\gamma^4-54\gamma^3+117\gamma^2-108\gamma+20)}+729\gamma^9-5382\gamma^8  \nonumber\\ &+&21384\gamma^7-49464\gamma^6+73872\gamma^5-57744\gamma^4
   -224\gamma^3+39168\gamma^2-28800 \gamma+6912.
   \end{eqnarray}
   Fig. \ref{frms} shows that both the marginally stable orbit radius $r_{\rm ms}$ and  the marginally bound orbit radius $r_{\rm mb}$ increase with the PPN parameter $\gamma$ of the CFM brane-world compact object. We also present the sizes of the singularity $r_{\rm s}$ and of the throat $r_{\rm throat}$ for different $\gamma$, and show that the marginally stable orbit and  the marginally bound orbit are outsides the naked singularity as $\gamma<1$ or the wormhole throat as $\gamma>1$.
   From Eq.(\ref{ELS}), one can find that the specific angular momentum and energy  $l$ for the particle moving along the circular orbit with the radius $r$ can be expressed as
    \begin{equation}
    \label{ell4}
    l^2\equiv\frac{L^2}{E^2}=\frac{ M \gamma^3 r^2}{r-2\gamma M+(\gamma-1)\sqrt{r(r-2\gamma M)}}\left(1-\frac{1}{\gamma}+\frac{1}{\gamma}\sqrt{1-\frac{2\gamma M}{r}} \right)^{-2},
    \end{equation}
   \begin{figure}
   \includegraphics[width=6cm]{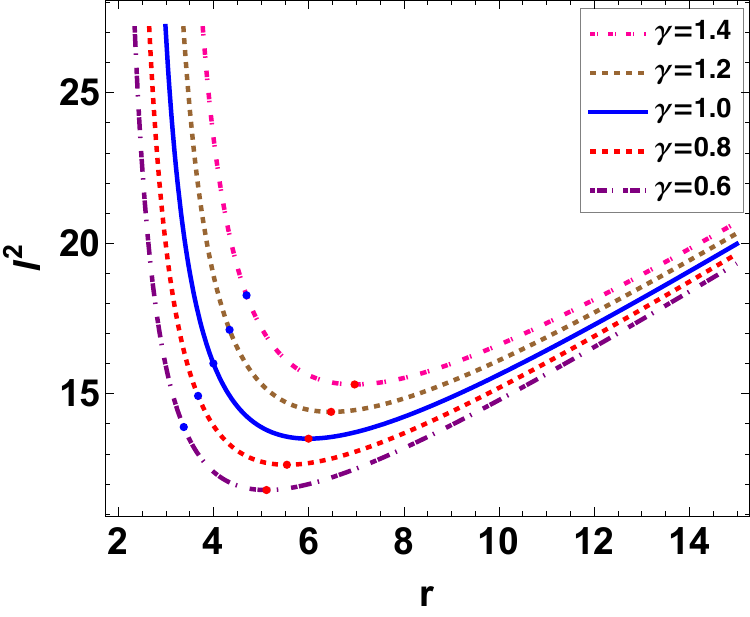}\;\;\;\includegraphics[width=6cm]{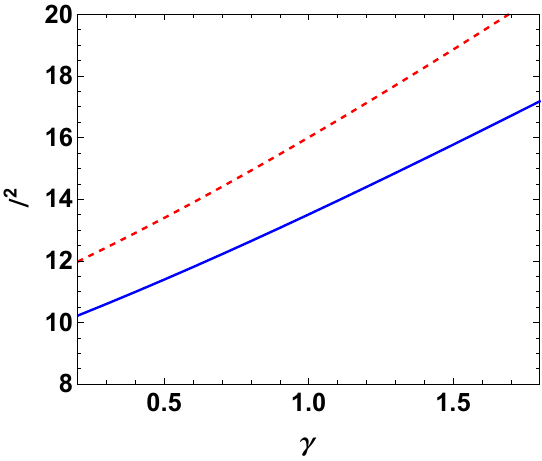}
   \caption{Changes of the specific angular momentum $l^2$ with the circular orbit radius $r$ for different PPN parameters $\gamma$. In the left panel, the blue and red dots  indicate the values of $l^2_{\rm mb}$ and $l^2_{\rm ms}$, respectively. In the right panel,
   the blue line and the dashed red line correspond  the values of $l^2$ at the marginally stable orbit $r_{\rm ms}$ and the marginally bound orbit $r_{\rm mb}$, respectively. Here we set $M=1$.}
   \label{llms}
   \end{figure}
 Above equations govern the particle motion. From Fig.\ref{llms}, one can find that the specific angular momentum for the particle moving along circular orbit increases with the spacetime parameter $\gamma$. Moreover, we also find that $l^2(r_{\rm mb})>l^2(r_{\rm ms})$ for each value of $\gamma$.

\section{Thick accretion disk configurations around the brane-world compact object}

Let us now to study thick accretion disk configurations around the CFM brane-world compact object (\ref{metric1}). As in previous works \cite{1975ApJ...199L.153E,1995ApJ...452..710N,Turolla_2000,Davis_2006,1977ApJ...214..840I,2002MNRAS.334..383F,
1982ApJ...253..897P,2022PhRvD.106h4046B,Cassing:2023bpt,Gimeno-Soler:2018pjd,Gimeno-Soler:2021ifv,Cruz-Osorio:2021gnz,Gimeno-Soler:2023anr}, here we adopt the test-fluid approximation where the accretion flow in the disk is a barotropic perfect fluid with positive pressure and its self-gravity is
negligible so that the influence of the disk on the background spacetime is negligible.
We also consider that the fluid is axisymmetric and stationary, which means that the physical variables
depend on only the coordinated $r$ and $\theta$. Finally, we assume that
the rotation of perfect fluid is restricted to be in the
azimuthal direction. With these assumptions, the four-velocity and stress-energy tensor of the perfect fluid can be expressed as
\cite{1978A&A....63..221A}
\begin{eqnarray}
u^{\mu}&=&(u^t,0,0,u^{\phi}),\quad\quad\quad
T_{\mu\nu}=(\epsilon + p) u_\mu u_\nu + p g_{\mu\nu},
\end{eqnarray}
where $\epsilon$ and $p$ are the total energy density and the pressure for a comoving observer, respectively.
The corresponding redshift factor in the static spacetime (\ref{metric1}) can be is given
by
\begin{eqnarray}
u_t=\sqrt{\frac{g_{tt}g_{\phi\phi}}{l^2g_{tt}+g_{\phi\phi}}},
\label{ut10}
\end{eqnarray}
where $l$ is the specific angular momentum.
From the conservation for the perfect fluid $\nabla_\nu T^\nu_{\ \, \mu} = 0$, one can obtain \cite{1978A&A....63..221A}
\begin{equation}
\label{p}
\frac{\nabla_{\mu} p}{\epsilon+p} = -\nabla_{\mu} \ln(u_t) + \frac{\Omega\nabla_{\mu}l}{1 - \Omega l},
\end{equation}
where $\Omega\equiv u^{\phi}/u^{t}$ is the angular velocity of the fluid. The specific angular momentum $l$ depends on the circular orbit radius of the particle motion and the covariant derivative $\nabla_{\mu}l$ describes the changes of the specific angular momentum $l$ for particles moving along two adjacent circular orbits in the fluid. For a barotropic fluid, $\epsilon$
is a function of $p$, so the the right-hand side of (\ref{p}) becomes a differential.  This implies that either $dl = 0$ or $\Omega = \Omega(l)$. This result is known
as the (relativistic) von Zeipel theorem.
For a barotropic fluid, one can get a solution of the above
equation by integration, i.e.,
\begin{equation}
\label{pps1}
\int^{p}_{p_{\rm in}}\frac{d p}{\epsilon+p} = -\ln\frac{|u_t|}{|(u_t)_{\rm in}|} + \int^{l}_{l_{\rm in}}\frac{\Omega dl}{1 - \Omega l}=W_{\rm in}-W.
\end{equation}
The subscript ``in" denotes that the quantity is evaluated at
at the inner edge of the disk. The potential $W$ determines the topologies of
equipotential surfaces in the disk. Therefore, once the expression $\Omega=\Omega(l)$ is given,
one can obtain the equipotential surfaces in the disk. Although in the real astrophysical
situations, $l$ would be given by certain dissipative processes with timescales
much longer than the dynamical timescale, such as the possible viscosity. It must be pointed out that  the viscosity in astrophysical accretion disks can not come from ordinary molecular viscosity since such ordinary viscosity are too weak to explain observed phenomena. Up to now, these dissipative processes are not yet fully understood. A possible alternative way of prescribing this
unknown dissipative processes is to directly set the angular momentum
$l$ in the model as a constant \cite{1978A&A....63..221A} or a non-constant angular
momentum distribution \cite{Lei:2008ui,Wielgus:2014nva}.  Here, we adopt the model with the constant distribution of
angular momentum $l=l_0$ to study the equilibrium configurations in the thick disk for different
parameters in the CFM brane-world compact object (\ref{metric1}). In this model, the  potential $W$ can be further simplified as
\begin{equation}
W =- \ln|u_t|.
\label{wut10}
\end{equation}
The thick disk configurations depend heavily on the specific angular momentum $l_0$. From Fig.\ref{llms}, one can obtain that there is a minimum $l_{\rm ms}$ for the specific angular momentum $l_0$, so the fluid with $l_0<l_{\rm ms}$ can not move along a circular orbit and it is no possible to exist a disk around a CFM brane-world compact object  in this case. When $l_0=l_{\rm ms}$, there exists only a ring around the compact object.  As $l_{\rm ms}<l_0<l_{\rm mb}$, it is found to exist bound disk structures with a cusp. As $l_0$ increases to $l_0 = l_{\rm mb}$, one can find that the cusp is located at the marginally closed surface that
just extents to infinity \cite{2022PhRvD.106h4046B,Cassing:2023bpt,Gimeno-Soler:2018pjd,Gimeno-Soler:2021ifv,Cruz-Osorio:2021gnz,Gimeno-Soler:2023anr}. As $l_0$ further increases to $l_0>l_{\rm mb}$, one can find that the disk still exists but not cusp.
\begin{figure}
   \includegraphics[width=4cm]{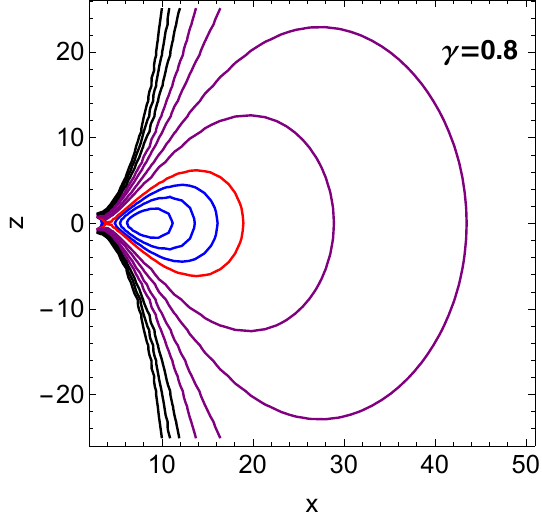};\; \includegraphics[width=4cm]{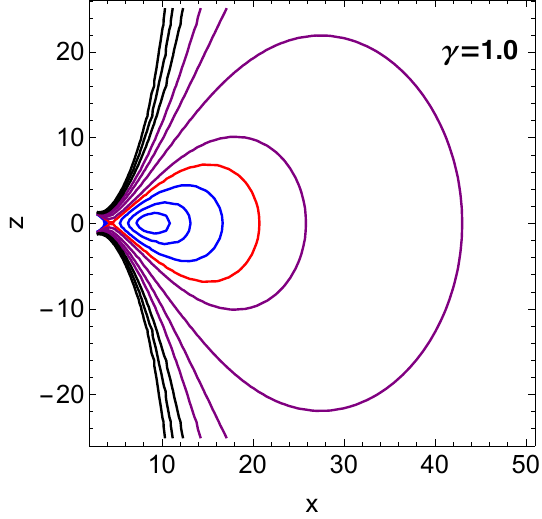}\;\;\includegraphics[width=4cm]{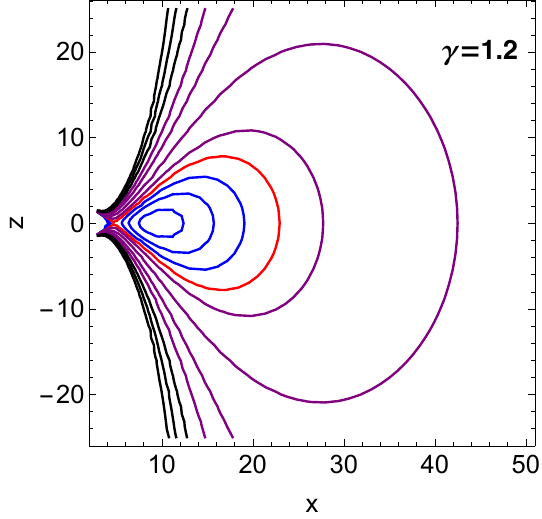}
    \caption{Equipotential surfaces for different choices of $\gamma$ and constant $l_0 = (l_{\rm mb}+l_{\rm ms})/2$. The red line indicates the torus with
a cusp corresponding to the maximum of $W$ on the equatorial plane. Blue lines indicate closed tori, purple lines bound structures
without inner edge, and black lines open surfaces.}
    \label{f5}
    \end{figure}
\begin{figure}
   \includegraphics[width=6cm]{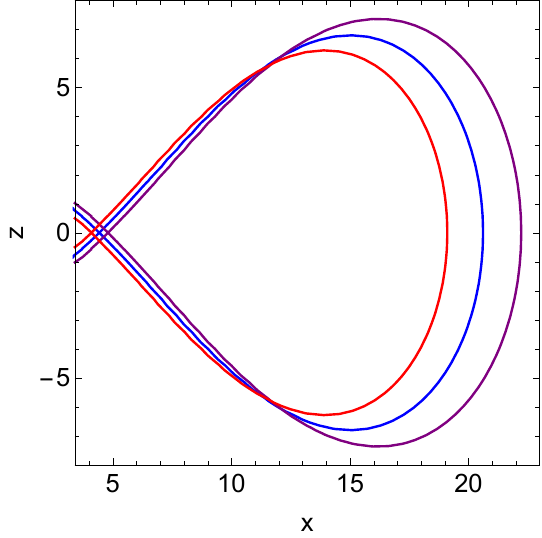}
    \caption{Equipotential surfaces corresponded to Roche lobe for different choices of $\gamma$ and constant $l_0 = (l_{\rm mb}+l_{\rm ms})/2$. The red, blue and purple lines denote the cases with $\gamma=0.8$, $1.0$ and $1.2$, respectively. }
    \label{f6}
    \end{figure}
\begin{figure}
   \includegraphics[width=6cm]{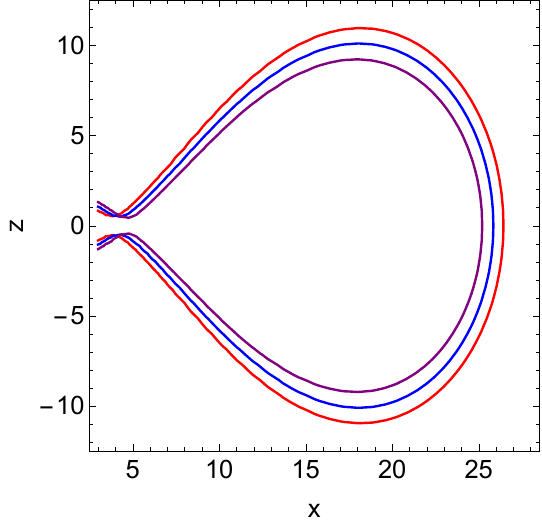}
    \caption{Equipotential surfaces with $W=-0.01$ outside the Roche lobe for different choices of $\gamma$ and constant $l_0 = (l_{\rm mb}+l_{\rm ms})/2$. The red, blue and purple lines denote the cases with $\gamma=0.8$, $1.0$ and $1.2$, respectively. }
    \label{f66}
    \end{figure}

 To probe the dependence of the disk configurations on the spacetime parameter $\gamma$, we set $l_0=\frac{1}{2}\left(l_{\rm ms}+l_{\rm mb}\right)$, and ensure $l_{\rm ms}<l_0<l_{\rm mb}$. Combining Eqs.(\ref{metric1}) with (\ref{ut10}) and (\ref{wut10}), one can obtain the potential function $W$ and probe the properties of the corresponding equipotential surfaces for different $\gamma$.
Fig. \ref{f5} shows the disk configurations around a CFM brane-world compact object  (\ref{metric1}).
In each panel, the blue lines denote possible bound disk
structures in which there are no actual accretion and the fluid rotate only around the compact object. The red line corresponds to the equipotential surface
with a cusp located at the marginally closed surface, which plays the same role as a Roche lobe and the matter from a disk outside
this surface would flow over the cusp and accrete
into the central compact object. The purple lines denote bound structures without inner edge but with a marginally outer edge,
and black lines denote the cases with open surfaces. The closed equipotential surface at infinite distance satisfies $W=0$.
With the increase of the parameter $\gamma$, we find that the value $W$ of the equipotential surface corresponded to the Roche lobe increases, and the surface of the Roche lobe gradually moves away from the central wormhole. The latter can be explained by a fact that both the marginally stable orbit radius $r_{\rm ms}$ and  the marginally bound orbit radius $r_{\rm mb}$ increase with the parameter $\gamma$ of the brane-world compact object  (\ref{metric1}). Moreover, with the increase of $\gamma$, the thickness of the region enclosed by the Roche lobe decreases near the compact object  and  increases for the region far from the compact object, but the area of the total region enclosed by the Roche lobe increases as shown in Fig. \ref{f6}, which means that the region of existing bound disk structures without accretion increases with the $\gamma$. From Fig. \ref{f6}, we also find that  the cusp is the point nearest the centre of the compact object in the Roche lobe. In the table (\ref{tabl1}), we compare the radial coordinate  of this cusp  $r_{\rm cusp}$  with the size of the compact object and find that the Roche lobe is outside the compact object.
\begin{table}[ht]
\centering
\begin{tabular}{|c|c|c|c|}\hline
$\gamma$ & 0.8 & 1.0& 1.2 \\ \hline
 $r_{\rm cusp}$&\quad4.082\quad&\quad4.437 \quad&\quad 4.809\quad\\ \hline
$r_{\rm s}$&\quad1.667&\;--& \;--  \\ \hline
$r_H $&--&\;2.0& \;--  \\
\hline
$r_{\rm throat}$&--&--& \;2.4  \\
\hline
\hline
\end{tabular}
\caption{Comparison between the cusp (the  nearest point to the centre of the compact object in the Roche lobe) and the size of the compact object for different $\gamma$. } \label{tabl1}
\end{table}
Fig. \ref{f66} presents the equipotential surface with $W=-0.01$ outside the Roche lobe for different $\gamma$, where the matter filling in this region can be accreted into the central compact object. It is illustrated that the size of accretion disk decreases with the PPN parameter of the CFM brane-world compact object, which means that the size of accretion disk for the CFM brane-world wormhole is less than that for the CFM brane-world naked singularity.

The pressure gradient in the thick disk is very important for the fluid to keep balance with gravitational and centrifugal
forces \cite{Cassing:2023bpt,Gimeno-Soler:2018pjd,Gimeno-Soler:2021ifv,Cruz-Osorio:2021gnz,Gimeno-Soler:2023anr}. Since the pressure gradient is related to the difference $\delta W$ between different equipotential surfaces,  we plot the maximal difference $\delta W$  between the potential values at the cusp and
the center of the disk.
\begin{figure}
   \includegraphics[width=6cm]{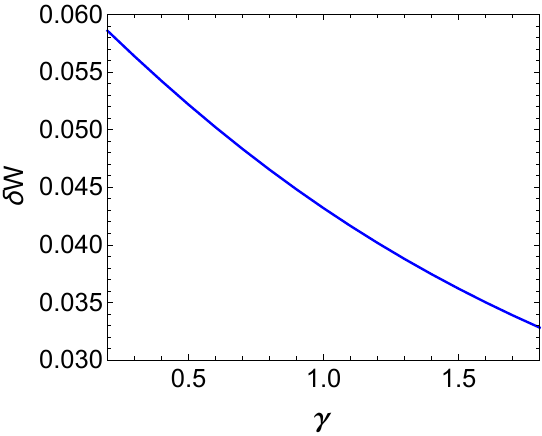}
    \caption{Change of $\delta W$ with the PPN parameter $\gamma$ for the brane-world compact object (\ref{metric1}) in the case $l_0=l_{\rm mb}$. }
    \label{f7}
\end{figure}
Fig. \ref{f7} illustrates that  the difference $\delta W$ and the pressure gradient in the equilibrium thick torus decrease with the PPN parameter $\gamma$ for the CFM brane-world compact object (\ref{metric1}).  It is also shown that the pressure gradient in the disk is larger in the CFM brane-world naked singularity, and is smaller in the CFM brane-world wormhole case.
These results could help to further understand compact objects in the brane-world.

\section{Summary}

We have studied the equipotential surfaces in the thick accretion disk around the CFM brane-world compact object with a PPN parameter. It is shown that with the increase of the PPN parameter, the size for the thick accretion disk decreases, but the Roche lobe increases. This implies that the larger PPN parameter results in the larger region of existing bound disk structures in which the fluid is not accreted into the central wormhole.
Moreover, with the increase of the parameter $\gamma$, the surface of the Roche lobe increases, and the Roche lobe gradually moves away from the central compact object. This can be explained by a fact that both the marginally stable orbit radius $r_{\rm ms}$ and  the marginally bound orbit radius $r_{\rm mb}$ increase with the parameter $\gamma$. In addition, the thickness of the region enclosed by the Roche lobe decreases with the parameter $\gamma$ near the compact object,  but increases in the region far from the compact object.  Finally, we have also studied the difference $\delta W$ between the potential values at the cusp and the center of the disk, which shows that the pressure gradient in the equilibrium thick torus the CFM brane-world compact object (\ref{metric1}) decreases with the parameter $\gamma$. Thus, the pressure gradient in the disk in the background of the CFM brane-world compact object is larger than that in the Schwarzschild background as $\gamma<1$, but is smaller as $\gamma>1$.   These results could help to understand the CFM brane-world compact object and its thick accretion disk.

\section{\bf Acknowledgments}
This work was  supported by the National Natural Science
Foundation of China under Grant No.12275078, 11875026, 12035005, and 2020YFC2201400.

    \bibliography{Brane-world-new}

    \end{document}